\documentclass[12pt]{iopart}
\usepackage{epsfig}
\usepackage{graphicx}

\newcommand{\beq}{\begin{equation}}
\newcommand{\eeq}{\end{equation}}
\newcommand{\bea}{\begin{eqnarray}}
\newcommand{\eea}{\end{eqnarray}}
\newcommand{\bce}{\begin{center}}
\newcommand{\ece}{\end{center}}

\def\lsim{\mathrel{\rlap{\lower4pt\hbox{\hskip1pt$\sim$}}
    \raise1pt\hbox{$<$}}}         
\def\gsim{\mathrel{\rlap{\lower4pt\hbox{\hskip1pt$\sim$}}
    \raise1pt\hbox{$>$}}}         

\begin{document}

\title{Theory and Phenomenology of Heavy Flavor at RHIC} 

\author{Ralf Rapp\footnote[3]{email: rapp@comp.tamu.edu} 
}

\address{Cyclotron Institute and Physics Department, Texas A\&M University, 
               College Station, Texas 77843-3366, U.S.A.}

\begin{abstract}
We review the problem of heavy-quark diffusion in the Quark-Gluon Plasma
and its ramifications for heavy-quark spectra in heavy-ion collisions at
RHIC. In particular, we attempt to reconcile underlying mechanisms of 
several seemingly different approaches that have been put forward to 
explain the large suppression and elliptic flow of non-photonic electron 
spectra. We also emphasize the
importance of a quantitative description of the bulk medium evolution to
extract reliable values for the heavy-quark diffusion coefficient. 
\end{abstract}




\section{Introduction}
\label{sec_intro}
The heavy-quark (HQ) mass, $m_Q\gg T_c$, brings a large scale into the 
problem of probing strongly interacting matter in the vicinity of its 
(pseudo-) critical temperature, $T_c$. This implies that in high-energy
heavy-ion collisions charm and bottom quarks ($Q$=$c$, $b$) are mostly
produced upon initial impact of the incoming nucleons, on a timescale 
$\tau_{\rm prod} \simeq 1/m_Q \le 0.1$~fm/$c$, while their abundance is 
expected to be frozen thereafter (which is supported by 
data~\cite{Adare:2006nq}). Therefore, changes in HQ momentum spectra 
due to reinteractions in the produced medium become an excellent (since 
rather direct) probe of the latter. The large HQ mass also implies that 
(the approach to) thermalization is delayed thus providing a more sensitive
means to study the in-medium interactions responsible for equilibration.
The (approximate) ``factorization" of $Q\bar Q$ production and rescattering
renders HQ observables a well-defined probe over the entire range of
transverse momentum ($p_T$), enabling a comprehensive study of 
(a) thermalization at low $p_T$, (b) a kinetic regime at intermediate
$p_T$ and (c) ``jet-quenching" at high $p_T$. The conservation of
individual charm and bottom quantum numbers allows to further test quark 
coalescence mechanisms in the hadronization transition. Theoretically, 
$m_Q\gg T$ opens the possibility to describe (low-$p_T$) HQ motion 
in the QGP within a diffusion equation~\cite{Svetitsky:1987gq}
which facilitates the extraction of pertinent transport 
coefficients from heavy-ion data. The transition from the elastic
to the radiative scattering regime is, of course, a key issue which
is closely related to items (a)-(c) above.
Finally, the availability of rather accurate lattice QCD (lQCD) computations
of HQ free energies at finite temperature has spurred the hope of
being able to define and extract interaction potentials which may not 
only govern the properties of heavy-quark ($Q\bar Q$) bound states but 
also heavy-quark diffusion. The nonperturbative nature of the latter 
has been exhibited in Refs.~\cite{CaronHuot:2007gq,vanHees:2007me}
(see Ref.~\cite{Rapp:2008qc} for a recent review).   

Our article is organized into a discussion of HQ diffusion coefficients
(Sec.\,\ref{sec_diff}), their applications to HQ spectra and 
observables at RHIC (Sec.\,\ref{sec_diff}) and conclusions
(Sec.\,\ref{sec_concl}).

\section{Heavy Quark Diffusion in the Quark-Gluon Plasma}
\label{sec_diff}
Brownian motion of a heavy quark in a fluid of light partons is 
described by a Fokker-Planck (FP) equation, schematically written as
\begin{equation}
\frac{\partial f_Q}{\partial t} = \gamma \frac{\partial pf_Q}{\partial p}
+ D \frac{\partial^2 f_Q}{\partial p^2} \ ,
\label{fp}
\end{equation}
which follows from the Boltzmann equation to 
second order in the momentum transfer, $k$, to the heavy quark. 
The friction ($\gamma$) and momentum diffusion ($D$) coefficients, 
\begin{equation}
\gamma p =\int d^3k \, w(k,p) \, k  \quad , \quad 
D=\frac{1}{2}\int d^3k \, w(k,p) \, k^2 \ , 
\end{equation}
are given in terms of transition rates, $w$, which are typically computed
from the  elastic scattering amplitude of the heavy quark with the (light) 
medium constituents. Both terms in Eq.~(\ref{fp}) are essential for the 
HQ distribution function, $f_Q$, to approach thermal equilibrium. 
This feature is highlighted by the Einstein relation, 
$T=D/(\gamma m_Q)$, which also shows that the FP equation involves only 
one independent transport coefficient. The latter is often quoted in 
terms of the spatial diffusion constant, $D_s=T/(\gamma m_Q)$.     

Early estimates of HQ diffusion in the QGP have utilized ``leading-order"
(LO) perturbative QCD (pQCD) which is dominated by $t$-channel gluon 
exchange with thermal partons (quarks and especially 
gluons)~\cite{Svetitsky:1987gq,vanHees:2004gq,Moore:2004tg,Mustafa:2004dr}. 
The infrared divergence for forward scattering is regulated by a thermal 
Debye mass, $\mu_D\sim gT$, resulting in a total cross section 
$\sigma_{Qp}\sim \alpha_s/\mu_D^2$. However, due to the forward-angle
dominated scattering, the transport cross section is much smaller; 
at temperatures around $T$=300\,MeV and for a strong coupling constant, 
$\alpha_s$=0.4, the thermal relaxation time for $c$-quarks amounts to 
$\tau=\gamma^{-1}\simeq$15-20\,fm/$c$ (larger by a factor of 
$\sim$$m_b/m_c$ for $b$ quarks). This is much longer than the expected 
QGP lifetime at RHIC and would therefore lead to small modifications of 
HQ spectra. 

\begin{figure}
\includegraphics[width=0.5\textwidth]{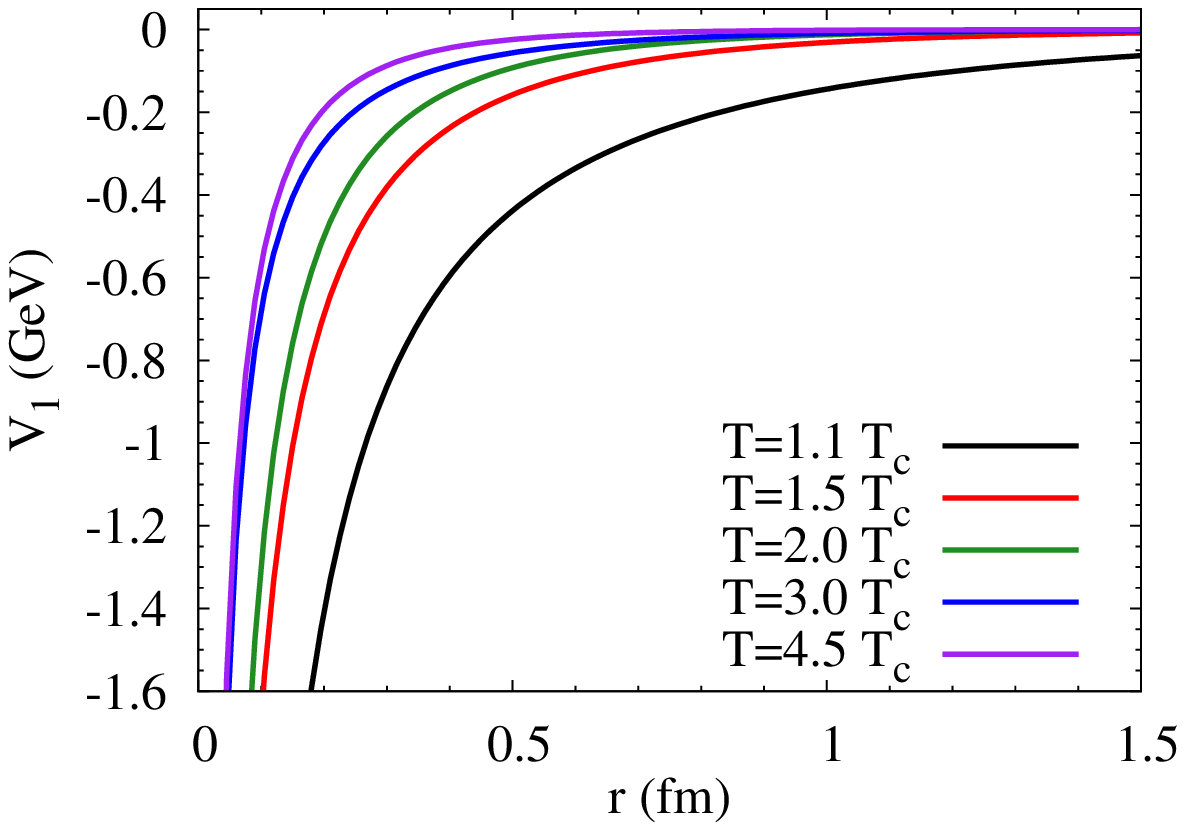}
\includegraphics[width=0.5\textwidth]{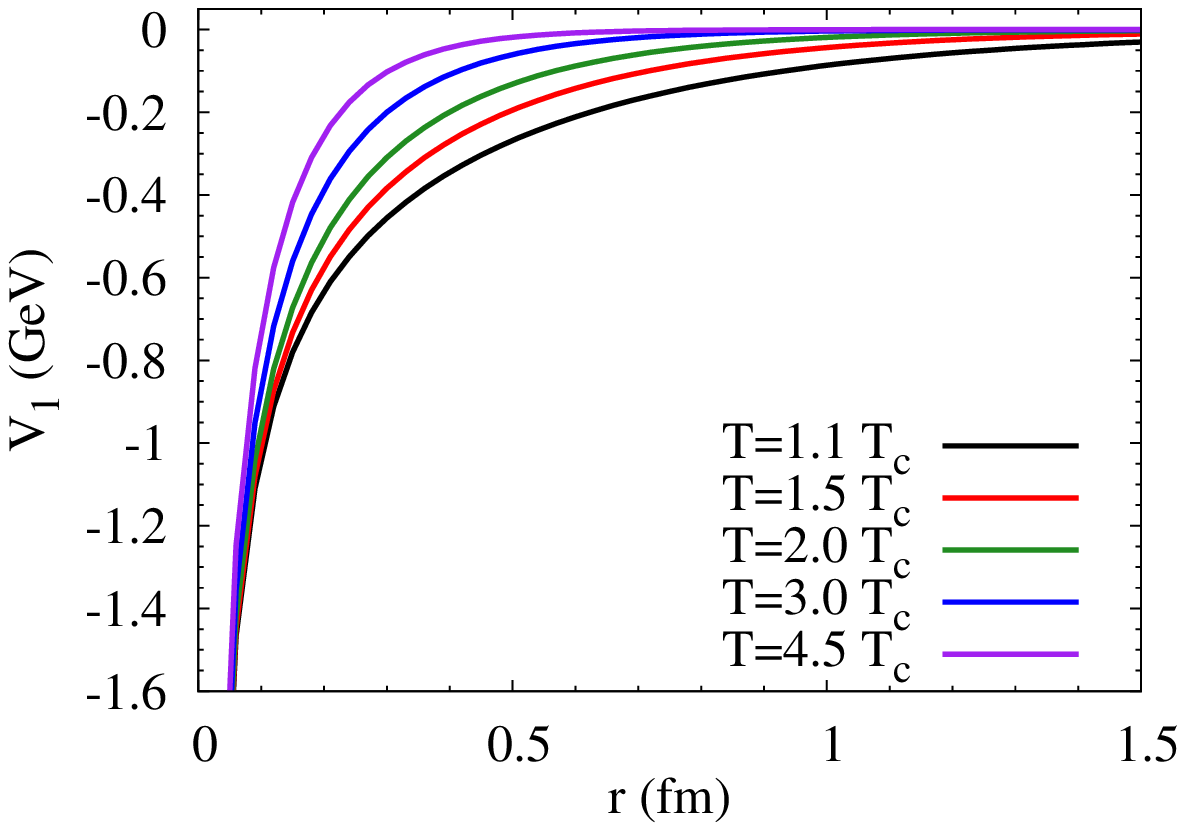}
\caption{Heavy-quark internal energies as extracted from 
fits~\cite{Wong:2004zr,Shuryak:2004tx} to quenched (left) and $N_f$=2 
(right) free energies computed in thermal 
lattice QCD~\cite{Kaczmarek:2003ph}.}
\label{fig_pot}
\end{figure}

In Ref.~\cite{vanHees:2004gq} it was suggested that mesonic resonances 
in the QGP could be operative in thermalizing heavy quarks. Pertinent 
effective Lagrangians where constructed utilizing HQ effective theory 
with a chirally symmetric set of $D$- and $B$-like resonances for 
$c$-$\bar q$ and $b$-$\bar q$ scattering. Within an estimated range 
of the underlying 2 parameters (coupling constant and resonance mass), 
the HQ relaxation times where found to be reduced by a factor $\sim$3 
compared to the LO-pQCD results. Pertinent Langevin simulations 
within an expanding QGP at RHIC, augmented with quark coalescence at 
$T_c$, resulted in a reasonable description~\cite{vanHees:2005wb} of 
non-photonic electron ($e^\pm$) observables at 
RHIC~\cite{Adare:2006nq,Abelev:2006db} 
(cf.~left panel of Fig.~\ref{fig_elec} below). The assumption of
resonances calls for a microscopic treatment to eliminate the free 
parameters of the effective Lagrangian and to assess their temperature 
dependence. In Ref.~\cite{vanHees:2007me}, a $T$-matrix equation for
heavy-light quark scattering was set up, which, after partial wave 
expansion ($L$=0,1,\dots), takes the form
\begin{equation}
\hspace{-1.0cm}
T_L^a(E_{cm};q,q') = V_L^a(q,q') + \int dk \, k^2 \, 
V_L^a(q,k) \, G_{Qq}(E_{cm},k) \, T_L^a(E_{cm};k,q') \ ; 
\label{Tmat}
\end{equation} 
$G_{Qq}$ denotes the intermediate 2-particle propagator which in 
principle contains in-medium selfenergies (mass corrections and width) 
of the individual heavy and light quark. The main idea is to implement
the interaction kernel, $V_L^a$, in a parameter-free way by utilizing
first-principle information from lQCD computations in terms
of the HQ free energy at finite temperature, 
\begin{equation}
F_{Q\bar Q}(T,r) = U_{Q\bar Q}(T,r) - T S_{Q\bar Q}(T,r) \ .
\end{equation} 
The same approach has been applied in recent years to compute heavy
quarkonium spectral functions in the QGP, with fair success in 
reproducing independent lQCD results for Euclidean-time correlation
functions~\cite{Mocsy:2007jz,Alberico:2007rg,Wong:2006bx,Cabrera:2006wh}. 
In the vacuum, $F_{Q\bar Q}(r)=U_{Q\bar Q}(r)$ closely resembles the 
phenomenological Cornell potential (Coulomb+confinement) which has been 
very successful in quarkonium spectroscopy and is now understood as the 
low-momentum limit of QCD with heavy quarks (so-called ``potential 
QCD")~\cite{Brambilla:2004wf}. The appropriate potential definition at 
finite temperature is currently an open question. An upper limit
may be obtained by using the (subtracted) internal energy, 
$V_{Q\bar Q}(r;T)$=$U_{Q\bar Q}(r;T)-U_{Q\bar Q}(r$=$\infty;T)$, which in 
the quarkonium sector provides the largest binding~\cite{Cabrera:2006wh}.
The subtraction is required to ensure convergence of the
Fourier transform of the potential in momentum space in Eq.~(\ref{Tmat}).  
Two examples of pertinent potential extractions are shown in 
Fig.~\ref{fig_pot}.
Their application in the $T$-matrix equation (\ref{Tmat}) additionally 
includes a relativistic correction to simulate color-magnetic
interactions~\cite{Shuryak:2004tx}. The Born 
approximation to the $T$-matrix, $T_L^a = V_L^a$, recovers the results 
from LO-pQCD within $\sim$10\% above $E_{cm}$=4~GeV 
(for the same value of $\alpha_s$). The different color channels,
$a$, are accounted for by Casimir scaling of the potentials. It turns
out that the resulting in-medium $T$-matrices are dominated by $S$-wave
scattering in the attractive color-singlet (meson) and -antitriplet 
(diquark) channels, supporting resonance-like structures close
the $qQ$ threshold up to $T$$\simeq$\,1.7\,$T_c$ and 1.4\,$T_c$, 
respectively.  
\begin{figure}
\begin{minipage}{0.5\textwidth}
\includegraphics[width=0.98\textwidth]{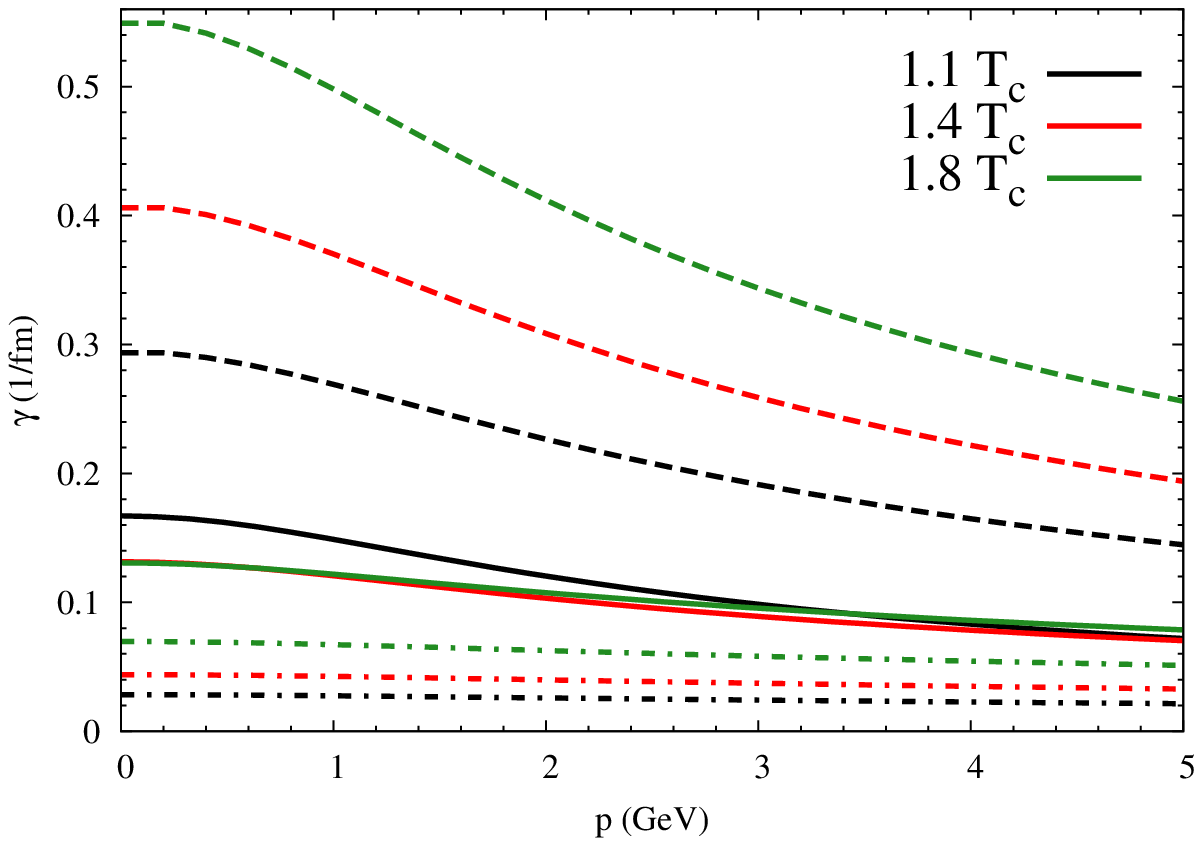}
\end{minipage}
\begin{minipage}{0.5\textwidth}
\includegraphics[width=0.98\textwidth]{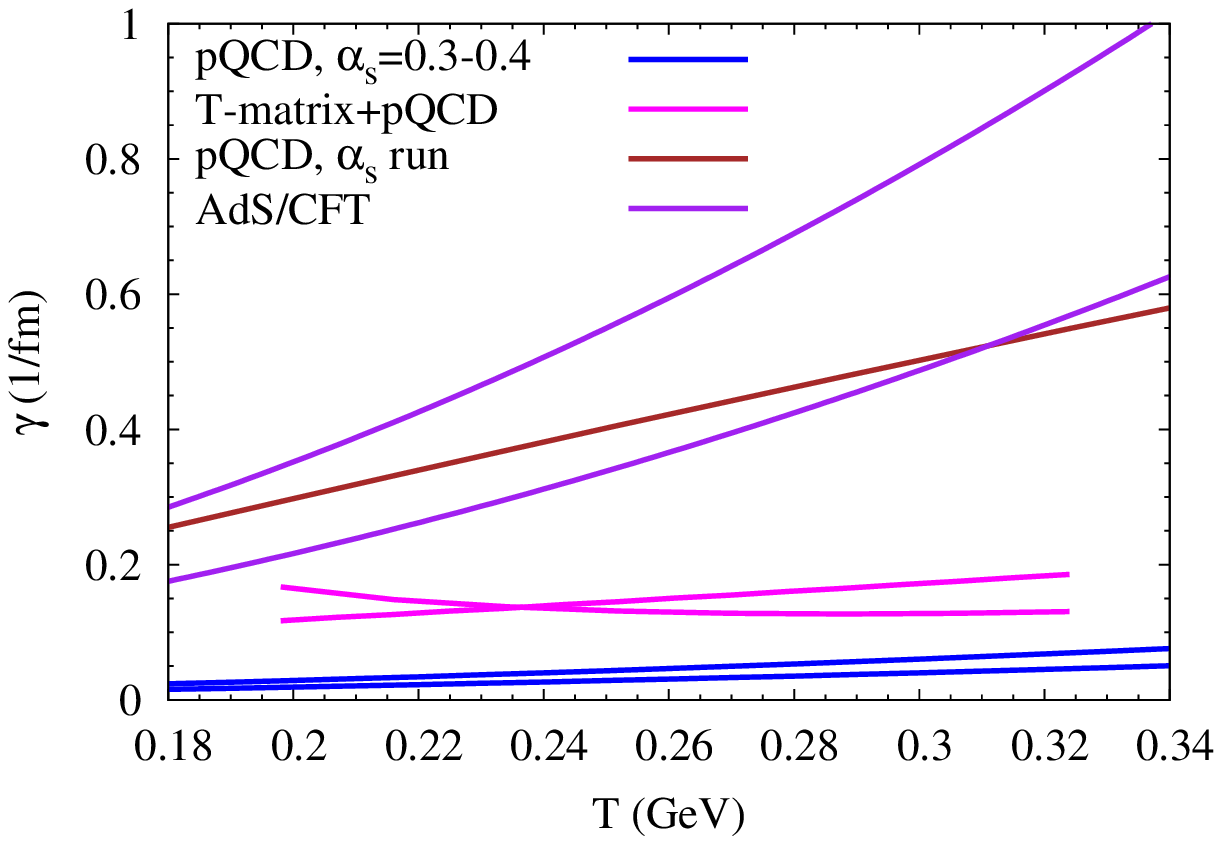}
\end{minipage}
\caption{Charm-quark friction coefficients $\gamma$ in the QGP. 
Left panel: 3-momentum dependence at 3 temperatures 
(color code) for: LO-pQCD with fixed $\alpha_s$=0.4 and $\mu_D$=$gT$ 
(lower 3 lines), heavy-light quark $T$-matrix plus LO-pQCD for gluons 
(middle 3 lines)~\cite{vanHees:2007me}, and pQCD with running $\alpha_s$ 
and reduced infrared regulator (upper 3 lines)~\cite{Gossiaux:2008jv}.
Right panel: temperature dependence of $\gamma$ for LO-pQCD, $T$-matrix
plus LO-pQCD (gluons only), pQCD with running $\alpha_s$, and from AdS/CFT
correspondence matched to QCD~\cite{Gubser:2006qh} with 
$C$=1.5-2.6~\cite{Akamatsu:2008ge} in
Eq.~(\ref{ads-cft}).} 
\label{fig_gam}
\end{figure}
Close to $T_c$, the corresponding friction coefficients are a factor
$\sim$3 larger than LO-pQCD (see Fig.~\ref{fig_gam}), but they 
{\em decrease} as temperature 
increases due to the weakening (color-screening) of the lQCD-based 
potential in the left panel of Fig.~\ref{fig_pot} (for the potential 
in the right panel of Fig.~\ref{fig_pot} a slight increase of 
$\gamma(T)$ is found).  
It is interesting to note that the collisional dissociation of $D$ and 
$B$ mesons in the QGP evaluated in Ref.~\cite{Adil:2006ra} is based on a 
similar Cornell-type potential as the $T$-matrix~\cite{vanHees:2007me}, 
but without the inclusion of medium
effects in the potential. The effect of a reduced formation time of $B$ 
relative to $D$ mesons~\cite{Adil:2006ra} is encoded in the $T$-matrix 
calculations as a mass effect leading to a stronger binding (analogous to 
the heavy quarkonium sector~\cite{Cabrera:2006wh}).  

Perturbative evaluations of HQ transport have recently been revisited in
Ref.~\cite{Gossiaux:2008jv}, where it is argued that the infrared regulator
in the $t$-channel gluon propagator becomes operative at a significantly 
softer scale than the Debye mass. In addition, a running of the strong 
coupling constant to small scales is implemented. 
The combined effect on the HQ friction coefficient is an approximately 
ten-fold increase over the Born result with Debye-mass 
regulator, cf.~Fig.~\ref{fig_gam}. Under these circumstances 
the perturbative treatment should be revisited and presumably augmented 
by resummations.

HQ diffusion has also been evaluated in conformal gauge theories 
where nonperturbative results can be inferred using a 
correspondence to string theory (AdS/CFT)~\cite{ads-cft}.
Defining the friction coefficient as the (inverse) timescale of 
momentum degradation, $dp/dt=-\gamma p$, cf.~Eq.(\ref{fp}), 
one finds for a $\cal{N}$=4 Super-Yang-Mills plasma: 
\begin{equation}
\gamma_{\rm AdS/CFT}=\frac{\pi \sqrt{\lambda} T_{\rm SYM}^2}{2m_Q}
\quad \leftrightarrow  \quad \gamma_{\rm QCD} = C \, \frac{T^2} {m_Q} \ . 
\label{ads-cft}
\end{equation}
The relation to a QCD plasma (second expression in Eq.~(\ref{ads-cft}))
requires a careful trans\-lation of the temperature scale and coupling 
constant; matching the energy densities of the SYM plasma and QGP to 
identify $T$ and using the static HQ force from lQCD to identify 
the coupling strength, $\lambda$, one finds~\cite{Gubser:2006qh}
$C$$\simeq$1.5-4. Note that this identification 
makes direct contact with the lQCD HQ free energy, as in the $T$-matrix 
approach.  

The right panel of Fig.~\ref{fig_gam} compiles the temperature dependence
of the friction coefficient, $\gamma$($p$=0), in the approaches discussed
above. The pQCD+running-$\alpha_s$ and AdS/CFT results are quite comparable
but significantly (much) larger than the ones from the $T$-matrix (LO-pQCD),
except close to $T_c$. However, it might be in the vicinity of $T_c$ where 
the QGP is most strongly coupled providing favorable conditions for the 
validity of the strong-coupling limit. The $T$-matrix approach exhibits the
weakest temperature dependence of the displayed curves, with a moderate
uncertainty induced by different version of the internal energy (however,
using the free energy as potential would give results closer to LO-pQCD). 

\begin{figure}
\begin{minipage}{0.5\textwidth}
\includegraphics[width=0.98\textwidth]{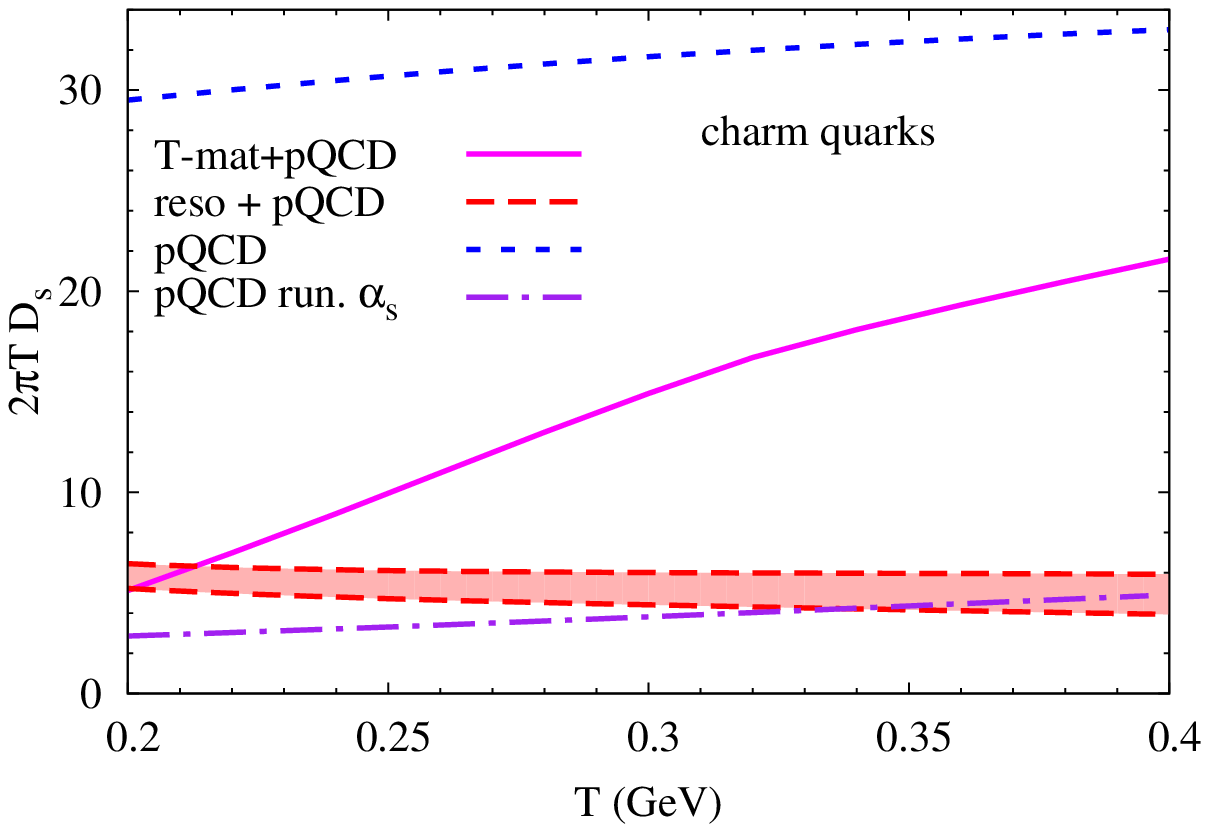}
\end{minipage}
\begin{minipage}{0.5\textwidth}
\includegraphics[width=0.98\textwidth]{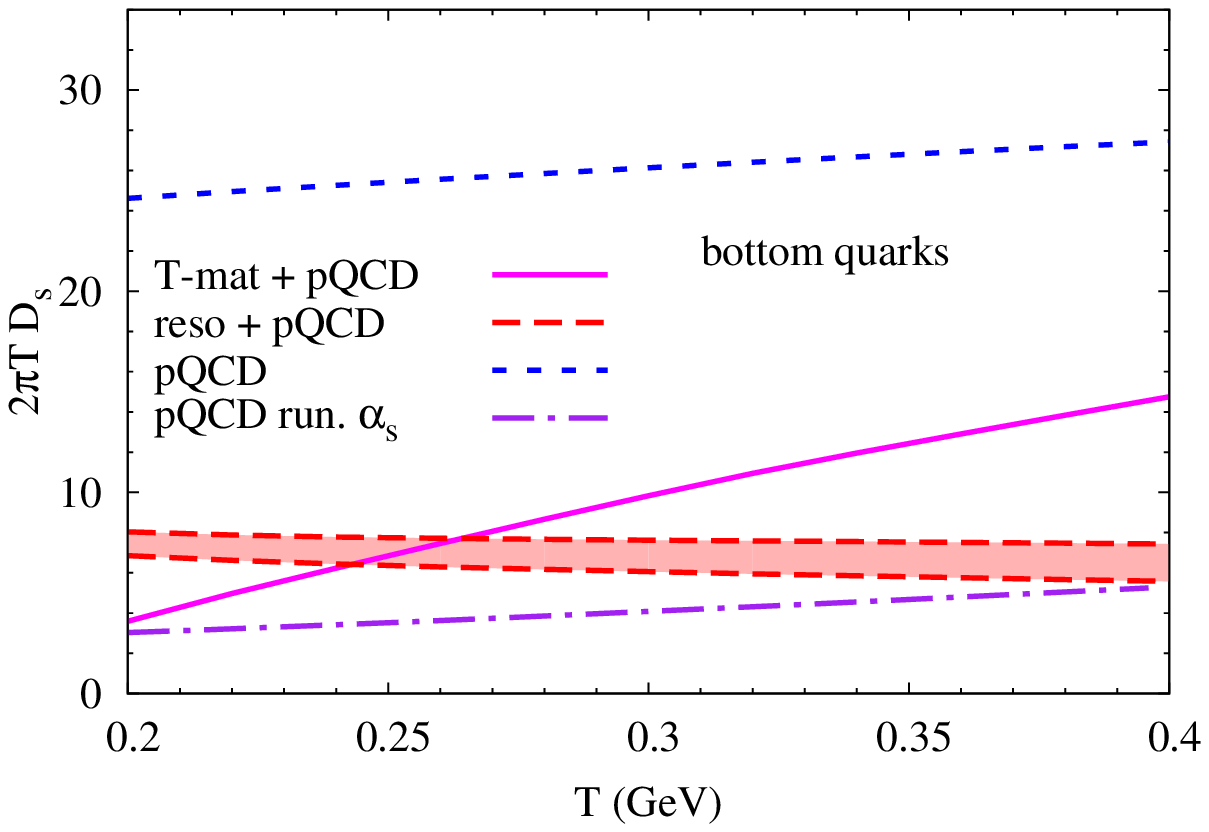}
\end{minipage}
\caption{Spatial diffusion coefficient, $D_s$=$T/(\gamma m_Q)$,
for $c$ (left) and $b$ quarks (right) in a QGP for:  
LO-pQCD with fixed $\alpha_s$=0.4
(dashed lines), effective resonance model + LO-pQCD (bands for 
$\Gamma_{D,B}$=0.4-0.75\,GeV)~\cite{vanHees:2004gq},  
$T$-matrix approach + LO-pQCD (gluons only)~\cite{vanHees:2007me}
and pQCD with running $\alpha_s$ (dash-dotted line)~\cite{Gossiaux:2008jv}.
The AdS/CFT result, Eq.~(\ref{ads-cft}), corresponds to 
$2\pi T D_s$=$2\pi/C$$\simeq$1.5-4 (not shown in the plots).}
\label{fig_Ds}
\end{figure}
Finally, we compare in Fig.~\ref{fig_Ds} the spatial diffusion 
coefficients (in units of the thermal Compton wavelength, $1/(2\pi T)$) 
at zero 3-momentum (which may be thought of as being proportional to 
the ratio of shear viscosity to entropy density) for charm and bottom 
quarks. All approaches give results fairly independent of temperature
and HQ mass, except for  the lQCD-based $T$-matrix calculation where
the increase with temperature indicates a significant loss of coupling
strength in the QGP.   

\section{Heavy-Flavor Spectra at RHIC}
\label{sec_rhic}
Several groups have applied the Brownian motion framework to simulate
HQ diffusion in Au-Au collisions at RHIC, using 
hydrodynamic~\cite{Moore:2004tg,Gossiaux:2008jv,Akamatsu:2008ge} or
expanding fireball~\cite{vanHees:2005wb} models.
Obviously, a realistic description of the bulk medium (temperature and 
flow evolution) is mandatory for a quantitative extraction of diffusion 
coefficients from data.

\begin{figure}
\includegraphics[width=0.5\textwidth]{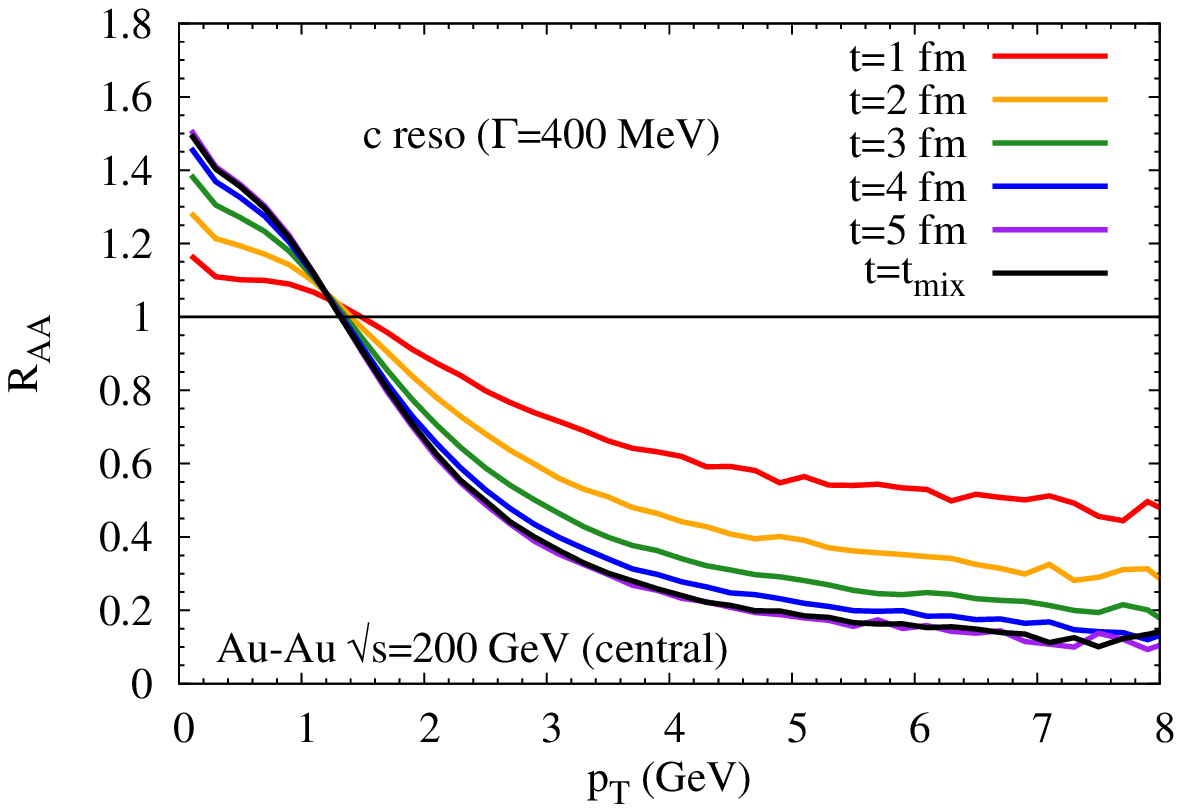}
\includegraphics[width=0.5\textwidth]{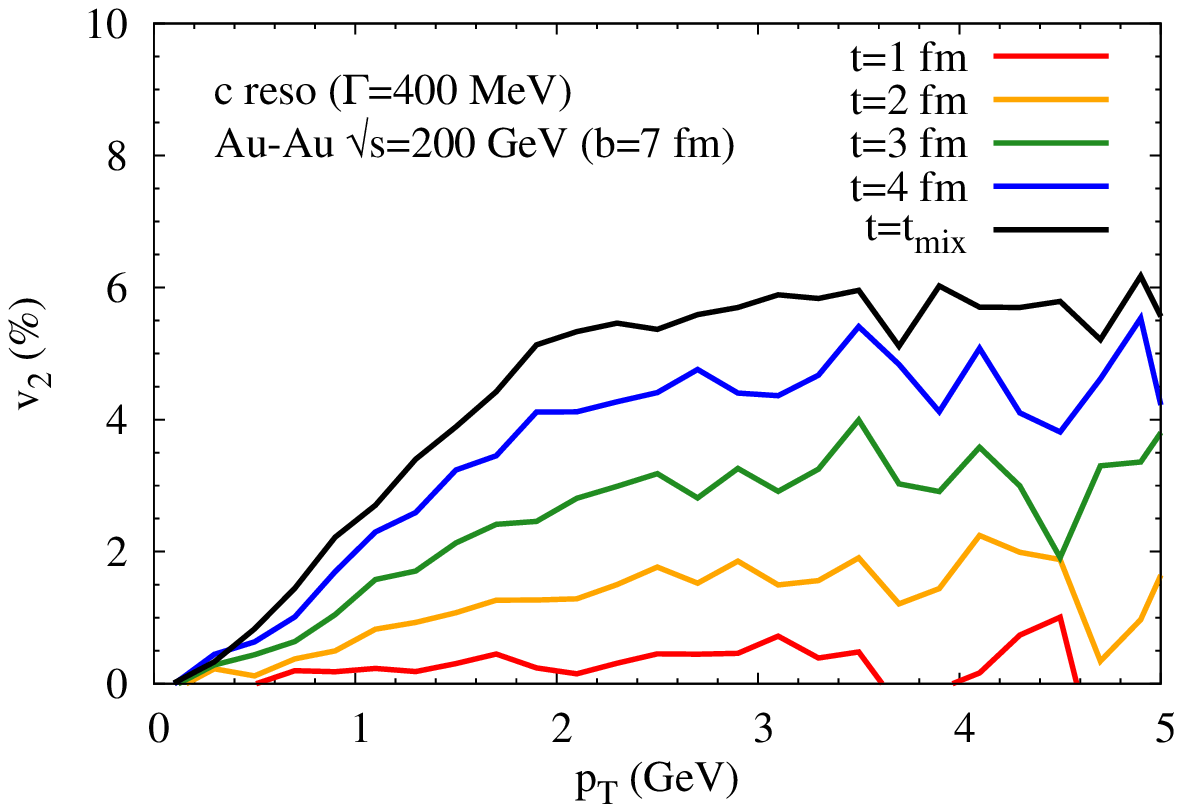}
\caption{Time evolution of nuclear modification factor 
(central collisions, left panel) and elliptic flow (semicentral 
collisions, right panel) in a QGP fireball at RHIC.~\cite{Rapp:2008qc}.}
\label{fig_time-evo}
\end{figure}

Let us first address the time evolution of HQ $p_T$-spectra, 
characterized by the nuclear modification factor 
$R_{AA}(p_T)=(dN_{AA}/dp_T)/(N_{\rm coll}\,dN_{pp}/dp_T)$,
and elliptic flow coefficient, $v_2(p_T)$. The results in 
Fig.~\ref{fig_time-evo}, computed in a thermal fireball evolution with
HQ transport based on resonance+LO-pQCD interactions~\cite{vanHees:2005wb}, 
suggest that the high-$p_T$ suppression is built up significantly earlier
than the elliptic flow. 
The former feature is quite reminiscent to a recent 
analysis~\cite{Wang:2008se} of the empirical system-size dependence 
of high-$p_T$ hadron suppression, arguing that parton energy-loss is
predominantly operative in the first 2-3\,fm/$c$.
On the contrary, the bulk $v_2$ in hydrodynamic models (used to construct 
the fireball evolution) requires a duration of at least 
$\Delta \tau$$\simeq$2-3\,fm/c to build up most of its strength, 
thus ``delaying" the transfer to heavy quarks.

\begin{table}[!b]
\begin{center}
\begin{tabular}{|c|c|c|c|c|}
\hline
Model [Ref.] & $D_s (2\pi T)$  
& $b$ [fm] & $v_2^{\rm max}$ & $R_{AA}$($p_T$=5~GeV) 
\\
\hline  \hline
hydro + LO-pQCD~\cite{Moore:2004tg} & 24  & 6.5 & 1.5\,\% & 0.7 \\
\hline
hydro + LO-pQCD~\cite{Moore:2004tg} & 6 & 6.5 & 5\,\% & 0.25 \\
\hline
fireball + LO-pQCD~\cite{vanHees:2005wb} & $\sim$30  & 7 & 2\,\% & 0.65 \\
\hline
fireball + reso+LO-pQCD~\cite{vanHees:2005wb} & $\sim$6 & 7 & 6\,\% & 0.3 \\
\hline
hydro + Eq.~(\ref{ads-cft})~\cite{Akamatsu:2008ge} 
& 21 & 7.1 & 1.5-2\,\% & $\sim$0.7 \\
\hline
hydro + Eq.~(\ref{ads-cft})~\cite{Akamatsu:2008ge} & 2$\pi$ & 7.1 & 4\,\% & $\sim$0.3 \\
\hline
\end{tabular}
\end{center}
\caption{Overview of model approaches (1.~column) and input parameters 
(2.~column: spatial charm-quark diffusion coefficient, 
3.~column: nuclear impact parameter) 
for Langevin simulations of charm-quark 
spectra in Au-Au collisions at RHIC; selected values for the resulting 
elliptic flow ($v_2^{\rm max}$$\simeq$$v_2$($p_T$=5\,GeV))
and nuclear modification factor are quoted in columns 4 and 5.}
\label{tab_hq}
\end{table}
Next we compare Langevin simulations for charm quarks in semicentral 
Au-Au collisions at RHIC. Representative values obtained for $v_2$ and 
$R_{AA}$ in 3 different evolution codes (2 hydro, 1 fireball), 
implementing 3 of the approaches discussed above for HQ transport 
coefficients, are summarized in Tab.~\ref{tab_hq}. Given the complete 
independence of the calculations, the agreement on the $\sim$30\,\% 
level is encouraging (it might improve when accounting for finer 
details, e.g., the results of Ref.~\cite{Akamatsu:2008ge} correspond
to terminating the evolution in the middle of the mixed phase while
in Refs.~\cite{Moore:2004tg,vanHees:2005wb} it is run to the
end of the mixed phase, cf.~right panel of Fig.~\ref{fig_time-evo}).

\begin{figure}
\begin{minipage}{0.5\textwidth}
\includegraphics[width=0.98\textwidth]{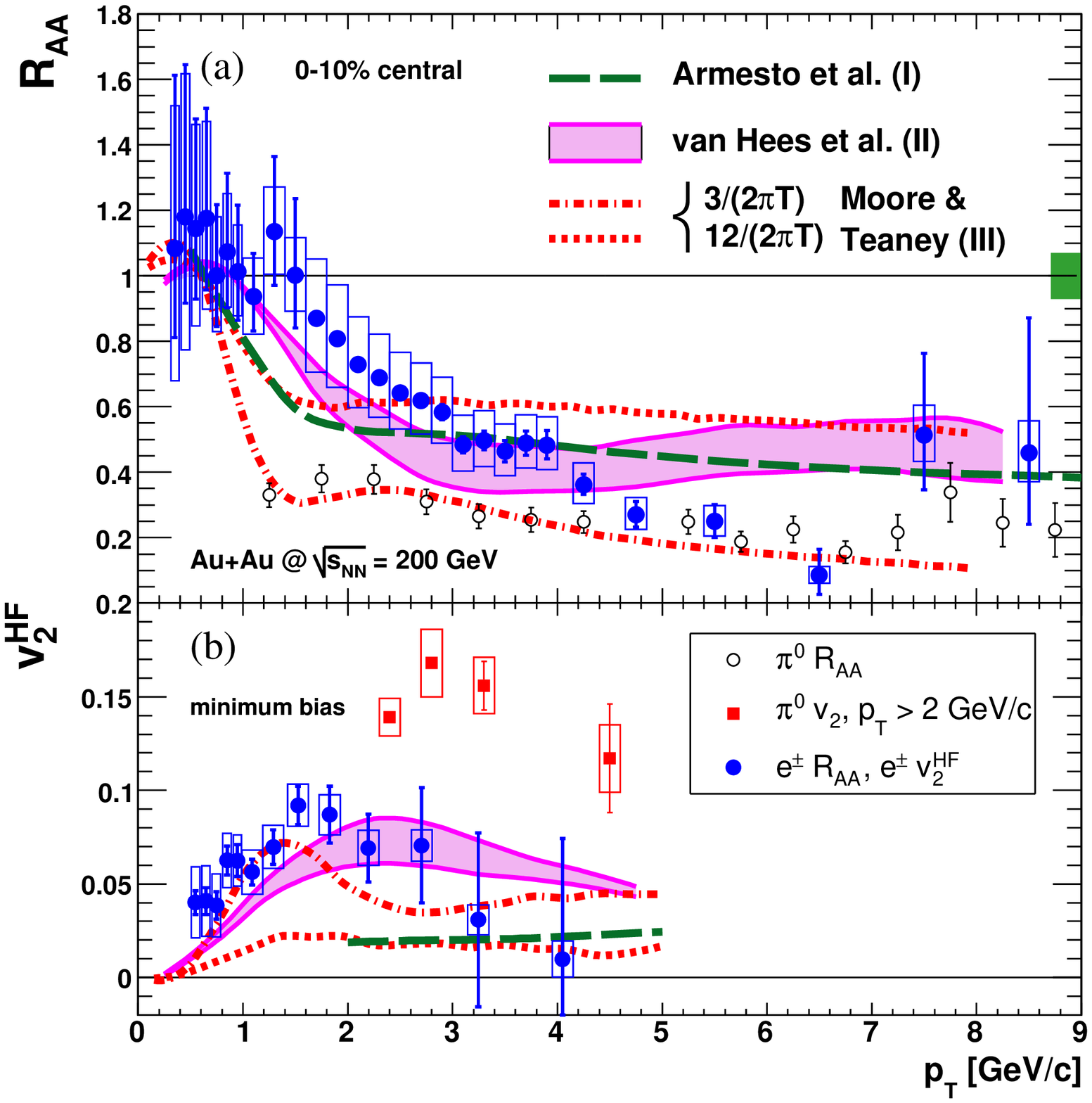}
\end{minipage}
\begin{minipage}{0.5\textwidth}
\vspace{0.45cm}
\includegraphics[width=0.98\textwidth]{RAA-v2-elec-tmat.eps}
\end{minipage}
\caption{Nuclear modification factor (upper panels, central Au-Au) and
elliptic flow (lower panels, minimum-bias Au-Au) of non-photonic 
electrons at RHIC~\cite{Adare:2006nq,Abelev:2006db,Hornback:2008ur}, 
compared to theory. Left panel~\cite{Adare:2006nq}: 
Langevin simulations based on 
(i) hydrodynamic evolution with (upscaled) LO-pQCD HQ interactions (dotted 
and dash-dotted lines)~\cite{Moore:2004tg} and (ii) expanding fireball 
with effective HQ resonance interactions (band)~\cite{vanHees:2005wb},
and pQCD radiative energy-loss calculations with (upscaled) transport 
coefficient in a static medium~\cite{Armesto:2005mz}; 
right panel: QGP fireball with $T$-matrix interactions~\cite{vanHees:2007me} 
using 2 extractions of the HQ internal energy 
from lattice QCD as potential (see Fig.~\ref{fig_pot}) where the dashed
lines do not include heavy-light quark coalescence at $T_c$.}
\label{fig_elec}
\end{figure}
Finally, theoretical predictions for non-photonic electron spectra 
are compared to RHIC data in Fig.~\ref{fig_elec}.
This requires the conversion
of the modified $c$- and $b$-quark spectra into meson spectra
followed by their semileptonic decay. For the radiative energy-loss 
calculations of Ref.~\cite{Armesto:2005mz} and the Langevin simulations 
of Ref.~\cite{Moore:2004tg} hadronization is treated via independent 
fragmentation, while in Refs.~\cite{vanHees:2005wb,vanHees:2007me}
coalescence processes with light quarks from the medium are accounted 
for. The latter enhance the hadron (and electron) $v_2$ while reducing
the suppression (see dashed lines in the right panel of 
Fig.~\ref{fig_elec}). While the radiative energy-loss calculations
(with upscaled pQCD transport coefficient) roughly account for the
observed suppression, the corresponding elliptic flow is underpredicted.
This underlines the importance of accounting for the collectivity of 
the expanding QGP medium, transferred to heavy quarks via the diffusion 
term in Eq.~(\ref{fp}). The current data-theory
comparison points at HQ diffusion coefficients in the range 
of $D_s(2\pi T)$$\simeq$4-6.

\section{Conclusions}
\label{sec_concl}
Several calculations of HQ diffusion in a QGP at moderate temperatures 
have been conducted recently but do not (yet?) show satisfactory 
agreement, neither in magnitude nor in their temperature dependence. 
We have argued, however, that there is significant overlap in the 
underlying physics, which in most cases is based on a (in-medium)
color-Coulomb type interaction as implicit in both one-gluon exchange 
and potential models (or in matching to conformal gauge theories). 
The challenge is thus to reconcile the different {\em approaches} 
(perturbative, $T$-matrix, AdS/CFT, etc.) by revisiting their 
regimes of applicability. The inclusion of radiative processes in a 
diffusion framework remains another challenge. On the empirical front, 
electron spectra at RHIC indicate a small HQ diffusion constant, but 
also here further scrutiny in the implementation of different medium 
evolution models (including the hadronic phase) needs to be exerted.

\vspace{0.5cm}

\noindent
{\bf Acknowledgment} \\
I thank the conference organizers for the invitation to a very nice 
meeting, and J.~Aiche\-lin, P.B.~Gossiaux and H.~van Hees for valuable 
discussions. 
This work was supported by a U.S.~National Science Foundation
CAREER award under grant no. PHY-0449489.

%

\end{document}